\documentclass[10pt]{article}
\usepackage{amsmath}
\usepackage{graphicx}
\usepackage{color}
\usepackage{orcidlink}
\usepackage{hyperref}
\hypersetup{colorlinks=true, linkcolor=blue, citecolor=blue, urlcolor=blue}
\usepackage{caption}
\usepackage{subcaption}
\begin{document}
\baselineskip=16pt

\begin{center}
{\large {\bf Scalar fields in Bonnor-Melvin-Lambda universe with potential: A study of dynamics of spin-zero particles-antiparticles }}
\end{center}

\vspace{0.3cm}

\begin{center}
    {\bf Faizuddin Ahmed\orcidlink{0000-0003-2196-9622}}\footnote{\bf faizuddinahmed15@gmail.com}\\
    \vspace{0.1cm}
    {\it Department of Physics, University of Science \& Technology Meghalaya, Ri-Bhoi, Meghalaya, 793101, India}\\
    \vspace{0.2cm}
    {\bf Abdelmalek Bouzenada\orcidlink{0000-0002-3363-980X}}\footnote{\bf abdelmalek.bouzenada@univ-tebessa.dz ; abdelmalekbouzenada@gmail.com (Corresp. author)}\\
    \vspace{0.1cm}
    {\it  Laboratory of theoretical and applied Physics, Echahid Cheikh Larbi Tebessi University, Algeria}\\
\end{center}

\vspace{0.3cm}

\begin{abstract}
In this study, our primary focus is on exploring the relativistic quantum dynamics of spin-zero scalar particles in a magnetic space-time background. Our investigation revolves around solving the Klein-Gordon (KG) equation within the framework of an electrovacuum space-time, while incorporating an external scalar potential. Specifically, we consider a cylindrical symmetric Bonnor-Melvin magnetic universe featuring a cosmological constant, where the magnetic field aligns parallel to the symmetry axis. Our approach involves deriving the radial equation of the wave equation, initially considering a linear confining potential and subsequently incorporating a Cornell-type scalar potential. We successfully obtain an approximate analytical solution for the eigenvalues of the quantum system under examination. Worth noting is our observation that the energy spectrum and the corresponding radial wave function experience notable modifications due to the presence of various factors including the cosmological constant, the topological parameter characterizing the space-time geometry, and the potential parameters.   
\end{abstract}

\vspace{0.1cm}

{\bf keywords}: Exact solutions: Bonnor-Melvin Universe; Relativistic Wave Equation: the Klein-Gordon equation; cosmological Constant; special functions; solutions of wave equations: bound-states. 

\vspace{0.1cm}


\section{Introduction }

Embarking on a profound exploration of the intricate interplay between gravitational field and the dynamics of quantum mechanical systems sparks significant interest. A. Einstein's groundbreaking general theory of relativity (GR) eloquently portrays gravity as an inherent geometric manifest of space-time \cite{k1}. This theory reveals the captivating connection between space-time curvature and the emergence of classical gravitational fields, providing precise predictions for phenomena such as gravitational waves \cite{k2, k10}, gravitational lensing \cite{k11, k12}, Lense-Thirring effect \cite{k8, k9}, and black holes \cite{k3}. Undoubtedly, one of the most captivating theories in physics is the general theory of relativity. On the other hand, quantum mechanics (QM) offers insights into the behaviors of particles at the microscopic scale \cite{k4} .

The convergence of these two realms of physics holds immense promise, offering a gateway to deeper insights into the fundamental nature of the universe. The success of quantum field theory in deciphering subatomic particle interactions and unraveling the origins of weak, strong, and electromagnetic forces \cite{k5} further heightens the anticipation surrounding this union. However, the longstanding pursuit of a unified theory-a theory of quantum gravity reconciling general relativity and quantum mechanics-has encountered persistent challenges and technical issues until very recently \cite{k6, k7}. These hurdles have spurred intense scientific endeavors as researchers diligently strive to bridge gaps in our understanding, aiming to unveil the foundational framework harmonizing these two essential cornerstones of modern physics.

In recent times, a notable interest has emerged in the study of magnetic field, driven by the discovery of systems with exceptionally strong field, such as magnetars \cite{k13, k14}, and occurrences in heavy ion collisions \cite{k15, k16, k17}. This raises a compelling question: how can the magnetic field be seamlessly integrated into the overarching framework of general relativity? This inquiry adds an intriguing dimension to the ongoing exploration of the universe's fundamental forces and their interconnected dynamics. Exploring the integration of the magnetic field within the framework of general relativity prompts intriguing questions. Significant strides have been made in comprehending the evolution of magnetic fields within the cosmological framework, particularly in exploring the mechanisms responsible for setting the fields to their current strengths in galaxies and galaxy clusters (for a comprehensive review, see \cite{LL1, LL2, LL3, LL4}). The models derived from the Einstein-Maxwell field equations are instrumental in elucidating various relativistic astrophysical phenomena. These phenomena encompass neutron stars characterized by intense gravitational fields, charged spheroidal stars, static charged and uncharged stars, cold compact objects, active galactic nuclei (AGN), and numerous others. Various exact solutions to the Einstein-Maxwell field equations were constructed, including the Manko solution \cite{k18, k19}, the Bonnor-Melvin universe \cite{k20, k21}, and a recently proposed solution \cite{MA} that incorporates the cosmological constant into the Bonnor-Melvin framework (see, also Refs. \cite{MA2, k22}). 

Transitioning our focus to the intersection of general relativity and quantum physics, a crucial consideration emerges: how these two theories may interrelate or if such a connection is even relevant. Numerous works have addressed this inquiry, predominantly relying on the Klein-Gordon and Dirac equations within curved space-times \cite{k23, k24}. This exploration extends to diverse scenarios, encompassing particles in Schwarzschild \cite{k25} and Kerr black holes \cite{k26}, cosmic string backgrounds \cite{k27, k28, k29}, quantum oscillators \cite{k30, k31, k32, k33, k34, k35, k35-1, k35-2, k35-3}, the Casimir effect \cite{k36, k37}, and particles within the Hartle-Thorne space-time \cite{k38}. These investigations, alongside others \cite{k39, k40, k41}, have yielded compelling insights into how quantum systems respond to the arbitrary geometries of space-time. Consequently, an intriguing avenue of study involves the examination of quantum particles within a space-time influenced by a magnetic field. For instance, in \cite{k42}, Dirac particles were explored in the Melvin metric, while our current work delves into the study of spin-0 bosons within the magnetic universe incorporating a cosmological constant, as proposed in \cite{k22}. The examination of generally covariant relativistic wave equations for a scalar particle in Riemannian space, characterized by the metric tensor $g_{\mu\nu}$ , necessitates the reformulation of the KG-equation given by as \cite{k35,k35-1,k35-2,k35-3,k43,k44,k45,ss1}
\begin{equation}
\Big[-\frac{1}{\sqrt{-g}}\,\partial_{\mu}\,\left(\sqrt{-g}\,g^{\mu\nu}\,\partial_{\nu}\right)+M^2\Big]\,\Psi ({\bf r})=0, \label{eq:1}
\end{equation}
where $M$ is the rest mass of the scalar particles, $g^{\mu\nu}$ is inverse of the metric tensor $g_{\mu\nu}$, and $\Psi ({\bf r})$ is the wave function.

We aim to study the relativistic quantum motions of spin-0 scalar particles in the background of an electrovacuum space-time. One of the example of such electrovacuum space-time is the Bonnor-Melvin magnetic universe, which is a generalization of Melvin universe, incorporating a cosmological constant. Furthermore, we introduce a static scalar potential by modifying the mass term in the Klein-Gordon wave equation and explore the dynamics of scalar fields. We derive the radial equation of the relativistic wave equation and solve it through special functions. We obtain the approximate eigenvalue solution of the quantum systems in the presence of a linear confining potential and a Cornell-type scalar potential. We show that the energy eigenvalue of the scalar fields are influenced by the cosmological constant, the topological parameter of the space-time, and the potential parameters. It is worth mentioning that quantum dynamics of scalar and oscillator field in the same Bonnor-Melvin universe that we consider here has recently been studied in Refs. \cite{k46,k47}.

This paper is summarized as follows: In {\it section 2}, we derive the radial wave equation of the Klein-Gordon equation in the presence of a non-electromagnetic scalar potential. We then solve this radial equation through special functions by considering two types of scalar potential: a linear confining and a Cornell-type potential and present the approximate energy spectrum and the wave function. In {\it section 3}, we present our results and discussion of the system under investigation. Throughout the paper we choose the system of units, where $\hbar=c=G=1$. 

\section{Scalar fields in Bonnor-Melvin-Lambda universe with potential: The Klein-Gordon Equation}

In this part, our focus turns to solve the Klein-Gordon equation within the background of Bonnor-Melvin space-time-enriched with a cosmological constant. The metric under consideration stands as static solution characterized by cylindrical symmetry, emerging from Einstein's field equations. This metric is intricately shaped by a homogeneous magnetic field, offering a compelling scenario for our investigation and the line-element describing this magnetic universe is given by \cite{MA, MA2, k22} 
\begin{equation}
ds^{2}=g_{\mu\nu}\,dx^{\mu}\,dx^{\nu}=-dt^{2}+d\rho^{2}+\sigma^{2}\,\sin^{2}(\sqrt{2\,\Lambda}\,\rho)\,d\varphi^{2}+dz^{2},\label{eq:3}
\end{equation}
where $\Lambda$ denotes the cosmological constant, and $\sigma$ represents a topological parameter which is responsible for an angular deficit. These two parameters are connected to the magnetic field strength along the symmetry axis through the following relation \cite{MA2, k22}:
\begin{equation}
H (\rho)=\sigma\,\sqrt{\Lambda}\,\sin\left(\sqrt{2\,\Lambda}\,\rho\right),\quad \Lambda>0.\label{eq:4}
\end{equation}

The behaviour of the magnetic field strength for different values of the cosmological constant $(\Lambda)$ and the topological parameter $(\alpha)$ is shown in the Figure 1.

\begin{center}
\begin{figure}
\subfloat[$\sigma=0.5$]{\centering{}\includegraphics[scale=0.5]{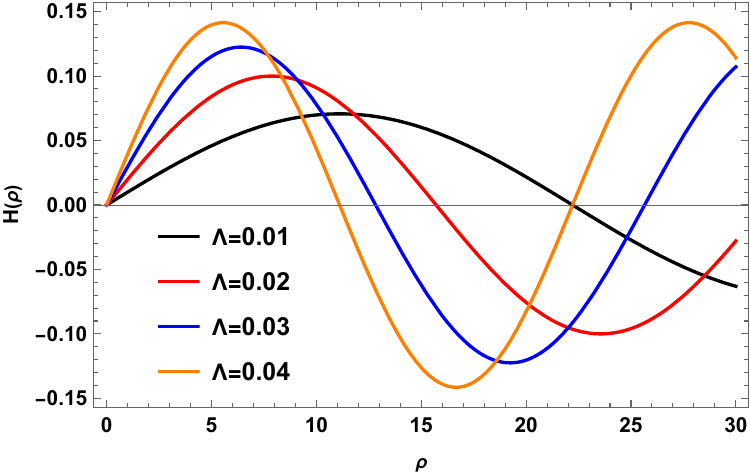}}\quad\quad\quad
\subfloat[$\Lambda=0.05$]{\centering{}\includegraphics[scale=0.5]{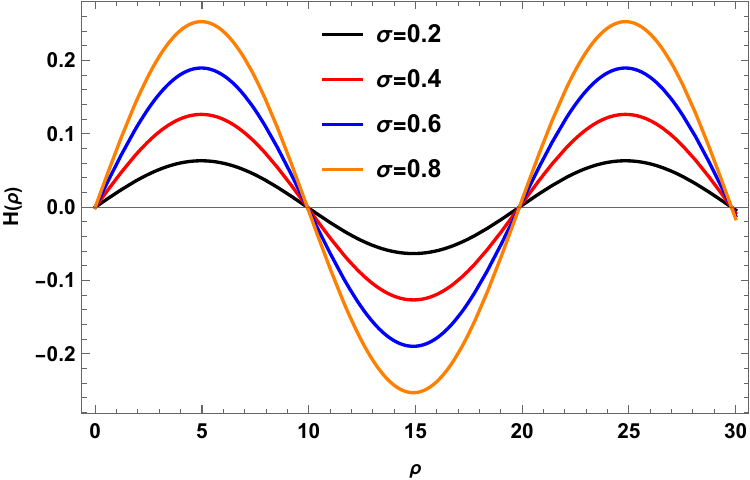}}\\
\centering
\subfloat[]{\centering{}\includegraphics[scale=0.5]{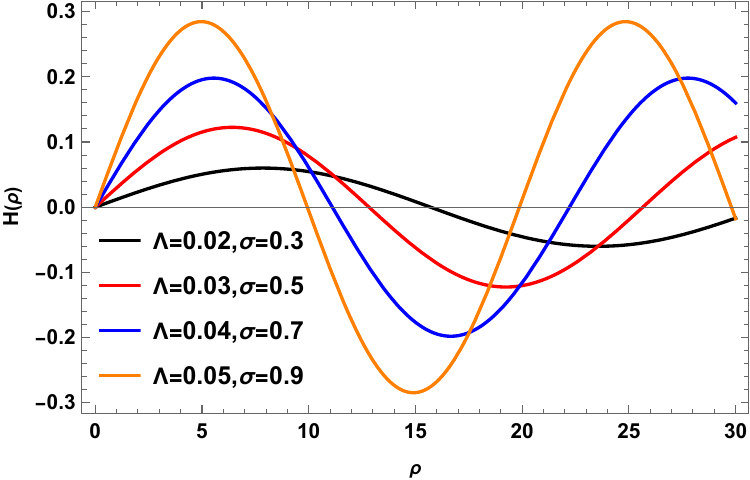}}
\caption{Variation of magnetic field strength with radial distance for different values of $\Lambda$ and $\sigma$.}
\end{figure}
\end{center}

The metric tensor $g_{\mu\nu}$ and it's inverse $g^{\mu\nu}$ for the line-element (\ref{eq:3}) are, respectively given by
\begin{equation}
g_{\mu\nu}=\left(\begin{array}{cccc}
-1 & 0 & 0 & 0\\
0 & 1 & 0 & 0\\
0 & 0 & \sigma^2\,\sin^2(\sqrt{2\,\Lambda}\,\rho) & 0\\
0 & 0 & 0 & 1
\end{array}\right),\quad 
g^{\mu\nu}=\left(\begin{array}{cccc}
-1 & 0 & 0 & 0\\
0 & 1 & 0 & 0\\
0 & 0 & \frac{1}{\sigma^2\,\sin^2(\sqrt{2\,\Lambda}\,\rho)} & 0\\
0 & 0 & 0 & 1
\end{array}\right).\label{eq:5}
\end{equation}

We introduce a static scalar potential into the Klein-Gordon equation (\ref{eq:1}) by modifying the mass term in this way: $M \to M+S(\rho)$, where $S$ is a position-dependent scalar potential \cite{WG}. Numerous authors have been studied the relativistic wave equations in the presence of scalar potentials of various kinds through this technique (see, Ref. \cite{k46}). Therefore, the modified KG-equation (\ref{eq:1}) in the presence of an external scalar potential becomes
\begin{equation}
\left[\frac{1}{\sqrt{-g}}\,\partial_{\mu}\,\left(\sqrt{-g}\,g^{\mu\nu}\,\partial_{\nu}\right)-(M+S)^2\right]\,\Psi=0. \label{eq:a1a}
\end{equation}
Expressing this wave equation (\ref{eq:a1a}) in the space-time background (\ref{eq:3}) and using (\ref{eq:5}), we obtain
\begin{eqnarray}
\Bigg[-\frac{\partial^{2}}{\partial t^{2}}+\frac{1}{\sin(\sqrt{2\,\Lambda}\,\rho)}\,\frac{\partial}{\partial \rho}\Big(\sin(\sqrt{2\,\Lambda}\,\rho)\, \frac{\partial}{\partial \rho} \Big)+\frac{1}{\sigma^2\,\sin^2 (\sqrt{2\,\Lambda}\,\rho)}\,\frac{\partial^2}{\partial \varphi^2}+\frac{\partial^{2}}{\partial z^{2}}-(M+S)^{2}\Bigg]\,\Psi=0,\label{eq:7}
\end{eqnarray}
where the determinant of the metric tensor $g_{\mu\nu}$ for the space-time (\ref{eq:3}) is given by 
\begin{equation}
\sqrt{-g}=\sigma\,\sin(\sqrt{2\,\Lambda}\,\rho).\label{eq:8}
\end{equation}

The differential equation (\ref{eq:7}) is independent of the time coordinate $t$, the angular coordinate $\varphi$, and the translation coordinate $z$. Therefore, we can choose one possible ansatz for the total wave function in the following form:
\begin{equation}
\Psi (t, \rho, \varphi, z)=e^{i\,(-E\,t+\ell\,\varphi+k\,z)}\,\psi(\rho),\label{eq:10}
\end{equation}
where $E$ is the particle's energy, $\ell$ is the angular momentum quantum number, and $k$ is an arbitrary constant.

Using this wave function (\ref{eq:10}) into the differential equation (\ref{eq:7}), we can write the following form of the second order differential equation 
\begin{equation}
\psi''+\frac{\sqrt{2\,\Lambda}}{\tan (\sqrt{2\,\Lambda}\,\rho)}\,\psi'+\Bigg[\beta^2-2\,M\,S(\rho)-S^2 (\rho)-\frac{\ell^2}{\sigma^2\,\sin^2 (\sqrt{2\,\Lambda}\,\rho)} \Bigg]\,\psi=0,\label{eq:11}
\end{equation}
where 
\begin{equation}
    \beta^2=E^2-M^2-k^2\,.\label{eq:12}
\end{equation}

We solve this radial equation by considering two different types of potential. Many potential models have been proposed in quantum system both in the context of relativistic and non-relativistic limit. Among these potential models are, linear confining potential, Yukawa potential, Coulomb potential which is used for H-atom problem, Cornell potential and many more. These potentials have wide applications including atomic and molecular physics, nuclear and particle physics.

\subsection{Linear potential}

In this part, we consider a linear potential into the above discussed quantum system and see how the relativistic scalar particles is influenced by this potential. Therefore, the linear confining potential is given by
\begin{equation}
    S=\eta\,\rho, \label{a1}
\end{equation}
where $\eta$ is the confining parameter. This potential has widely been used in the confinement of quark phenomenon \cite{CLC}, and in the quantum system by numerous authors (see, Refs. \cite{ERFM,FA,ZW}).

Thereby, substituting this confining potential into the radial equation (\ref{eq:11}) results the following equation:
\begin{equation}
\psi''+\frac{\sqrt{2\,\Lambda}}{\tan (\sqrt{2\,\Lambda}\,\rho)}\,\psi'+\Bigg[\beta^2-2\,M\,\eta\,\rho-\eta^2\,\rho^2-\frac{\ell^2}{\sigma^2\,\sin^2 (\sqrt{2\,\Lambda}\,\rho)} \Bigg]\,\psi=0,\label{eq:a2}
\end{equation}

The exact solution of the above differential equation is very difficult and complicated. Therefore, to obtain an eigenvalue solution, we assume that the cosmological constant $\Lambda$ is very small, and therefore, taking approximation up to the first order, we arrive from (\ref{eq:a2}) the following differential equation form: 
\begin{equation}
\psi''+\frac{1}{\rho}\,\psi'+\Bigg[\beta^2-2\,M\,\eta\,\rho-\eta^2\,\rho^2-\frac{\tau^2}{\rho^2} \Bigg]\,\psi=0,\label{a3}
\end{equation}
where $\tau=\frac{|\ell|}{\sigma\,\sqrt{2\,\Lambda}}$.

Now, we choose the following ansatz for the above differential equation:
\begin{equation}
\psi(\rho)=\rho^{\tau}\,\mathrm{e}^{-\frac{1}{2}\,\rho^{2}\,\eta-M\,\rho}\,\chi(\rho)\label{eq:a3}
\end{equation}
which is regular everywhere including at $\rho \to 0$.

Thereby, substituting this wave function in Eq. (\ref{a3}), we obtain the following differential equation form:
\begin{equation}
\chi''+\left[\frac{(2\,\tau+1)}{\rho}-2\,\eta\,\rho-2\,M\right]\,\chi'+\left[\beta^{2}+M^{2}-2\,(\tau+1)\,\eta-\frac{M\,(2\,\tau+1)}{\rho}\right]\chi=0.\label{eq:a4}
\end{equation}
Finally transforming to a new variable via $s=\sqrt{\eta}\,\rho$ in Eq. (\ref{eq:a4}) results the following form:
\begin{eqnarray}
s\,\chi''(s)+\left(1+2\,\tau-\alpha'\,s-2\,s^2\right)\,\chi' (s)+\left(\mu\,s-\nu\right)\,\chi(s)=0, \label{eq:a6}
\end{eqnarray}
where we set the parameters
\begin{equation}
\alpha'=\frac{2\,M}{\sqrt{\eta}},\quad \mu=\frac{\beta^{2}+M^{2}}{\eta}-2\,(1+\tau),\quad \nu=\frac{M}{\sqrt{\eta}}\,(1+2\,\tau).\label{eq:a8}
\end{equation}

Equation (\ref{eq:a6}) is the Biconfluent Heun second-order differential equation form \cite{AR} whose solution is well-known, and thus, $\chi (s)$ is the biconfluent Heun function given by the following form:
\begin{equation}
\chi (s)=\mathrm{HeunB}\left(2\,\tau, \alpha', \mu+2\,(1+\tau), 0, s\right).\label{eq:a9}
\end{equation}
Now, we solve the differential equation (\ref{eq:a6}) using a power series solution method. Therefore, we consider the following power series around the origin given by\cite{GBA}
\begin{equation}
\chi=\sum_{j=0}^{\infty}\mathcal{C}_{j}\,s^{j}.\label{eq:a10}
\end{equation}
Therefore, we obtain
\begin{equation}
\chi'=\sum_{j=1}^{\infty}\,j\,\mathcal{C}_{j}\,s^{j-1}=\sum_{j=0}^{\infty}\,(j+1)\,\mathcal{C}_{j+1}\,s^{j}.\label{eq:a11}
\end{equation}
And its second-order derivative gives
\begin{equation}
\chi''(s)=\sum_{j=2}^{\infty}\,j\,\left(j-1\right)\mathcal{C}_{j}s^{j-2}=\sum_{j=0}^{\infty}\,(j+1)\,(j+2)\,\mathcal{C}_{j+2}\,s^{j}.\label{eq:a12}
\end{equation}

Thereby, substituting equations (\ref{eq:a10})--(\ref{eq:a12}) in equation (\ref{eq:a6}), we get the following three-terms recurrence relation given by
\begin{equation}
\mathcal{C}_{j+2}=\frac{\left(\alpha'\left(j+1\right)+\nu\right)}{\left[\left(j+2\right)\left(2\left|\tau\right|+j+2\right)\right]}\mathcal{C}_{j+1}+\frac{\left(2j-\mu\right)}{\left[\left(j+2\right)\left(2\left|\tau\right|+j+2\right)\right]}\mathcal{C}_{j},\label{eq:aa17}
\end{equation}
with the following coefficient
\begin{equation}
\mathcal{C}_{1}=\frac{\nu}{2\,\tau+1}\mathcal{C}_{0}.\label{eq:a16}
\end{equation}

To obtain bound-state solution of the quantum system under investigation, we must truncate the three-terms recurrence relation such that the power series function (\ref{eq:a10}) becomes a finite degree polynomial. This truncation of a power series is obtained by setting the conditions $\mathcal{C}_{j+1}=0$ and $\mu=2\,j$ in the recurrence relation (\ref{eq:a17}) results the higher power coefficients of $s$ vanishes, that is, $\mathcal{C}_{j+2}=0$ and so on. In that case, the power series becomes $\chi=\sum_{j=0}^{n}\,\mathcal{C}_{j}\,s^{j}$ becomes a finite degree polynomial of degree $j$, and thus, the wave function (\ref{eq:a3}) is finite and well-behaved everywhere. The truncating conditions are given by
\begin{equation}
    \mathcal{C}_{j+1}=0,\quad \mu=2\,j.\label{eq:a17}
\end{equation}

Simplifying the second condition gives us the approximate (but not exact) energy eigenvalue expression associated with the mode $\{j, \ell\}$ given by
\begin{equation}
    E_{j,\ell}=\pm\,\sqrt{k^2+2\,\eta\,(j+1+\tau)}.\label{eq:a18}
\end{equation}
The radial wave function is given by
\begin{equation}
\psi_{j,\ell}=\mathcal{N}_{1}\rho^{\tau}\mathrm{e}^{-\frac{1}{2}\,\rho^{2}\,\eta-M\,\rho}\,\mathrm{HeunB}\,\Big(2\,\tau,\frac{2\,M}{\sqrt{\eta}},\frac{M^{2}+\beta^{2}}{\eta},0,\sqrt{\eta}\,\rho\Big).\label{eq:a21}
\end{equation}

One should remembered that the first condition given by $\mathcal{C}_{j+1}=0$ must also analyze simultaneously to get a complete information of the quantum system under investigation. This analysis will give us the individual energy level expression and the corresponding radial function of the scalar particles. The lowest state of the quantum system is defined by $j=1$. In that case, the coefficient $\mathcal{C}_2=0$. Therefore, the ground-state approximate energy level from (\ref{eq:a18}) will be 
\begin{equation}
    E_{1,\ell}=\pm\,\sqrt{k^2+2\,\eta\,\Big(2+\frac{|\ell|}{\sigma\,\sqrt{2\,\Lambda}}\Big)}.\label{eq:a19}
\end{equation}
From the recurrence relation (\ref{eq:a17}), we obtain
\begin{equation}
    \mathcal{C}_2=\frac{1}{2\,(1+2\,\tau)}\Bigg[(\alpha'+\nu)\,\mathcal{C}_1-\mu\,\mathcal{C}_0\Bigg].\label{eq:a20}
\end{equation}

For the ground state we have $\mathcal{C}_2=0$ and $\mu=2$ which implies $\mathcal{C}_1=\frac{2}{(\alpha'+\nu)}\,\mathcal{C}_0$. Therefore, equating this one with the equation (\ref{eq:a16}) results the following condition
\begin{equation}
    \eta_{1,\ell}=\Big(\frac{3}{2}+\frac{|\ell|}{\sigma\,\sqrt{2\,\Lambda}}\Big)\,M^2\label{aa}
\end{equation}
a constraint on the confining potential parameter $\eta \to \eta_{1,\ell}$ that permit us to construct a first degree polynomial of $\chi (\rho)$. We see that this parameter $\eta_{1,\ell}$ depends on the quantum number $\ell$ and the geometric parameters $(\sigma, \Lambda)$. 

Therefore, from equation (\ref{eq:a19}) we obtain the final expression of approximate energy level of the ground-state given by
\begin{equation}
    E_{1,\ell}=\pm\,\sqrt{k^2+2\,M^2\,\Big(\frac{3}{2}+\frac{|\ell|}{\sigma\,\sqrt{2\,\Lambda}}\Big)\Big(2+\frac{|\ell|}{\sigma\,\sqrt{2\,\Lambda}}\Big)},\quad \Lambda>0.\label{eq:aa19}
\end{equation}
And the corresponding approximate ground state wave function will be
\begin{equation}
    \psi_{1,\ell}=\rho^{\frac{|\ell|}{\sigma\,\sqrt{2\,\Lambda}}}\,\mathrm{e}^{-\frac{1}{2}\,\Big(\frac{3}{2}+\frac{|\ell|}{\sigma\,\sqrt{2\,\Lambda}}\Big)\,M^2\,\rho^{2}-M\,\rho}\,\Bigg[1+\Big(\frac{3}{2}+\frac{|\ell|}{\sigma\,\sqrt{2\,\Lambda}}\Big)\,\rho\Bigg]\,\mathcal{C}_0.\label{eq:aa3}
\end{equation}

Equation (\ref{eq:aa19}) is the relativistic approximate energy level and equation (\ref{eq:aa3}) is the corresponding wave function associated with ground state of the system in the background of Bonnor-Melvin magnetic universe in the presence of a linear confining scalar potential. Following the similar procedure, we one find excited states energy levels and wave function of the quantum system defined by the mode $j \geq 2$. We see that the ground-state energy spectrum and the wave function is influenced by the topological parameter $\sigma$ of the geometry which produces an angular deficit in the space-time, and the cosmological constant $\Lambda$. We also see that the ground-state approximate energy level $E_{1,\ell}$ for particle-antiparticles are equally spaced on either side about $E=0$ for a fixed quantum number $\ell$, indicates that energies are same both for particles-antiparticles.

\subsection{Cornell-type potential}

In this part, we consider a linear plus Coulomb-type potential together called Cornell-type potential given by
\begin{equation}
    S=\eta\,\rho+\frac{\xi}{\rho}, \label{b1}
\end{equation}
where $\xi>0$ is potential parameter. The linear part is responsible for long range interactions whereas the Coulomb part for short range interactions. The Cornell potential consists of a linear potential plus a Coulomb potential and a harmonic term that is used in quark-antiquark interaction \cite{MKB}. This type of potential has widely been used in the studies of quantum mechanical system by numerous authors (see, Refs. \cite{k41, ZW, ERFM}. 

Thereby, substituting this Cornell-type potential into the radial equation (\ref{eq:11}) and considering first order approximation (assuming the cosmological constant is very small) reduces to the following differential equation form:
\begin{equation}
\psi''+\frac{1}{\rho}\,\psi'+\Bigg[\beta^2-2\,\eta\,\xi-2\,M\,\eta\,\rho-\eta^2\,\rho^2-\frac{2\,M\,\xi}{\rho}-\frac{\iota^2}{\rho^2} \Bigg]\,\psi=0,\label{b2}
\end{equation}
where 
\begin{equation}
\iota^2=\xi^2+\tau^2=\xi^2+\frac{\ell^2/\sigma^2}{2\,\Lambda}.\label{b3}    
\end{equation}

Analogue to the previous section, let us consider the following wave function ansatz:
\begin{equation}
\psi\left(\rho\right)=\rho^{\left|\iota\right|}\mathrm{e}^{-\frac{1}{2}\rho^{2}\eta-M\rho}\Omega(\rho).\label{eq:b3}
\end{equation}
Substituting this wave function into the equation (\ref{b2}) results the following differential equation form:
\begin{equation}
\Omega''+\left[-2\,M-2\,\eta\,\rho+\frac{(2\,\iota+1)}{\rho}\right]\,\Omega'+\left[M^{2}+\beta^{2}-2\,\xi\,\eta-2\,(1+\iota)\,\eta-\frac{2\,M}{\rho}\left(\xi+\iota+\frac{1}{2}\right)\right]\,\Omega=0.\label{eq:b4}
\end{equation}
Transforming to a new variable via the following transformation $\mathcal{P}=\sqrt{\eta}\rho$ into the above differential equation results
\begin{equation}
\Omega''+\left[\frac{2\,\iota+1}{\mathcal{P}}-\frac{2\,M}{\sqrt{\eta}}-2\,\mathcal{P}\right]\Omega'+\left[\frac{M^{2}+\beta^{2}-2\,\xi\,\eta}{\eta}-2\,(\iota+1)-\frac{\left(\frac{M}{\sqrt{\eta}}\right)}{\mathcal{P}}\,(2\,\xi+2\,\iota+1)\right]\,\Omega=0.\label{eq:b6}
\end{equation}
Defining the following parameters
\begin{equation}
\delta=\frac{2M}{\sqrt{\eta}},\quad \gamma=\frac{M^{2}+\beta^{2}-2\,\xi\,\eta}{\eta}-2\,(\iota+1),\quad \varsigma=\frac{M}{\sqrt{\eta}}(2\,\xi+2\,\iota+1)\label{eq:b7}
\end{equation}
into the equation (\ref{eq:b6}), we obtain the following standard differential equation form given by
\begin{equation}
\mathcal{P}\,\Omega''+\left(1+2\,\iota-\delta\,\mathcal{P}-2\,\mathcal{P}^{2}\right)\,\Omega'+\left(\gamma\,\mathcal{P}-\varsigma\right)\,\Omega=0.\label{eq:b8}
\end{equation}
This differential equation is the Biconfluent Heun equation form \cite{AR} whose solution is given by
\begin{equation}
\Omega\left(\mathcal{P}\right)=\mathrm{HeunB}\left(2\,\iota,\delta,\gamma+2\,(\iota+1),2\,\varsigma-4\,(\iota+1),\mathcal{P}\right).\label{eq:b9}
\end{equation}

Following the previous analysis, we choose a power series given by \cite{GBA}
\begin{equation}
\Omega\left(\mathcal{P}\right)=\sum_{j=0}^{\infty}\mathcal{D}_{j}\mathcal{P}^{j}\label{eq:b10}
\end{equation}
into the differential equation (\ref{eq:b8}), we get the following coefficient 
\begin{equation}
\mathcal{D}_{1}=\frac{\varsigma}{\left[(2\,\iota+1)\right]}\,\mathcal{D}_{0}\label{eq:b14}
\end{equation}
with the three-terms recurrence relation given by
\begin{equation}
\mathcal{D}_{j+2}=\frac{\left[\alpha'\,(j+1)+\varsigma\right]}{\left[\left(j+2\right)\left(2\left|\iota\right|+j+2\right)\right]}\mathcal{D}_{j+1}+\frac{\left[2j-\gamma\right]}{\left[\left(j+2\right)\left(2\left|\iota\right|+j+2\right)\right]}\mathcal{D}_{j}.\label{eq:b16}
\end{equation}
The truncating conditions of the power series are given by
\begin{equation}
    \mathcal{D}_{j+1}=0,\quad \gamma=2\,j.\label{eq:b17}
\end{equation}
Simplification of the second condition gives the following expression of approximate energy eigenvalues given by 
\begin{equation}
    E_{j,\ell}=\pm\,\sqrt{k^2+2\,\eta\,\Bigg(j+1+\xi+\sqrt{\xi^2+\frac{\ell^2}{2\,\Lambda\,\sigma^2}}\Bigg)}.\label{eq:b18}
\end{equation}

\begin{center}
\begin{figure}
\centering
\subfloat[]{\centering{}\includegraphics[scale=0.5]{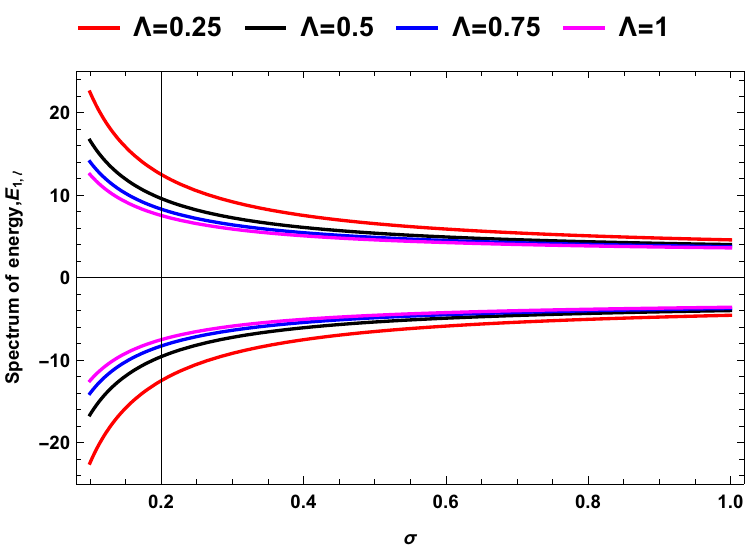}}\quad\quad\quad
\subfloat[]{\centering{}\includegraphics[scale=0.5]{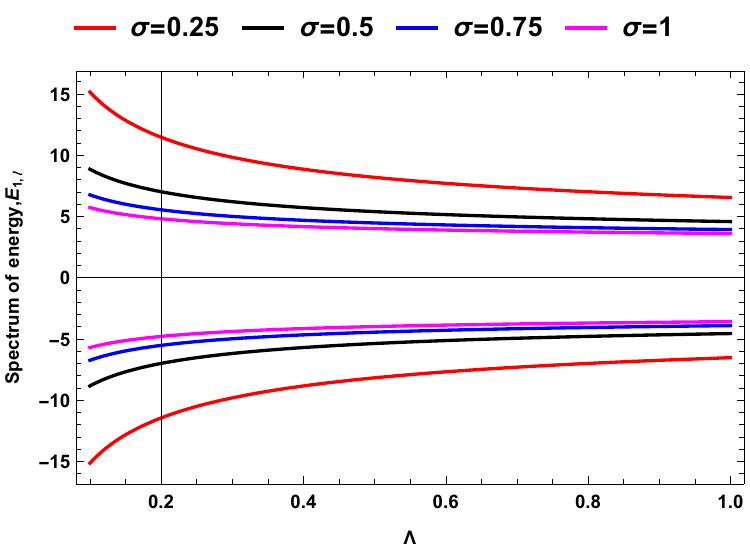}}\caption{Energy spectrum with $\sigma$ and $\Lambda$. Here $k=\ell=M=1$.}
\subfloat[$\Lambda=0.5$]{\centering{}\includegraphics[scale=0.5]{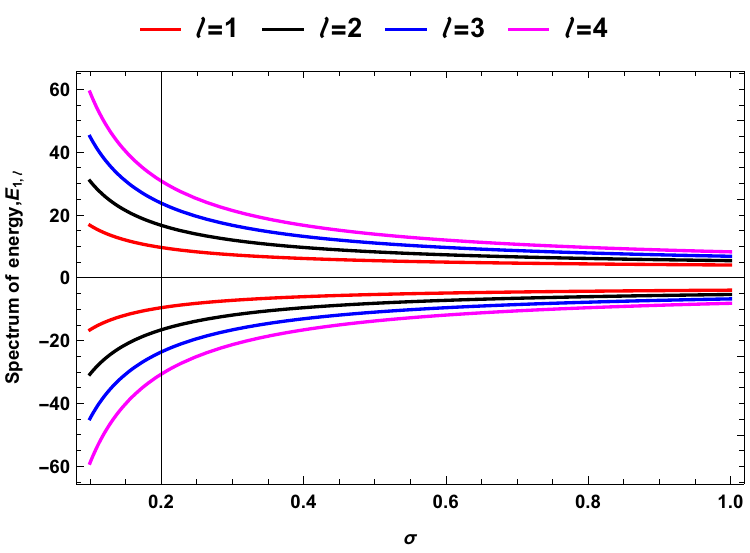}}\quad\quad\quad
\subfloat[$\sigma=0.5$]{\centering{}\includegraphics[scale=0.5]{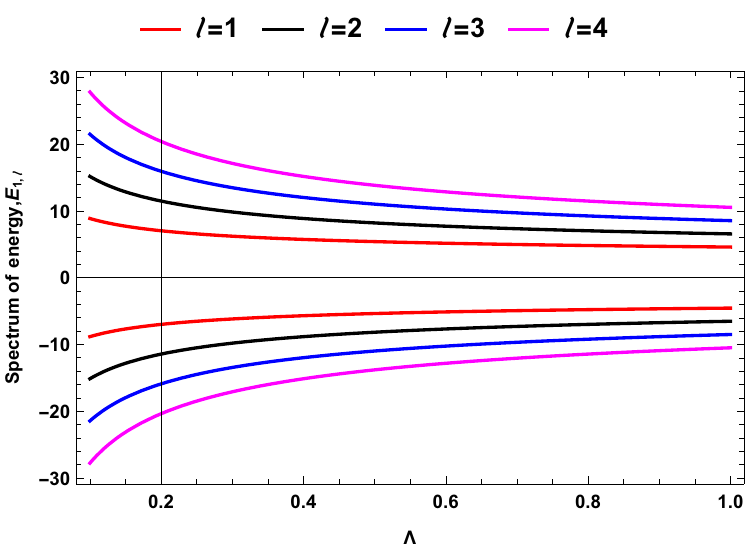}}\caption{Energy spectrum with $\sigma$ and $\Lambda$. Here $k=1=M$.}
\subfloat[]{\centering{}\includegraphics[scale=0.5]{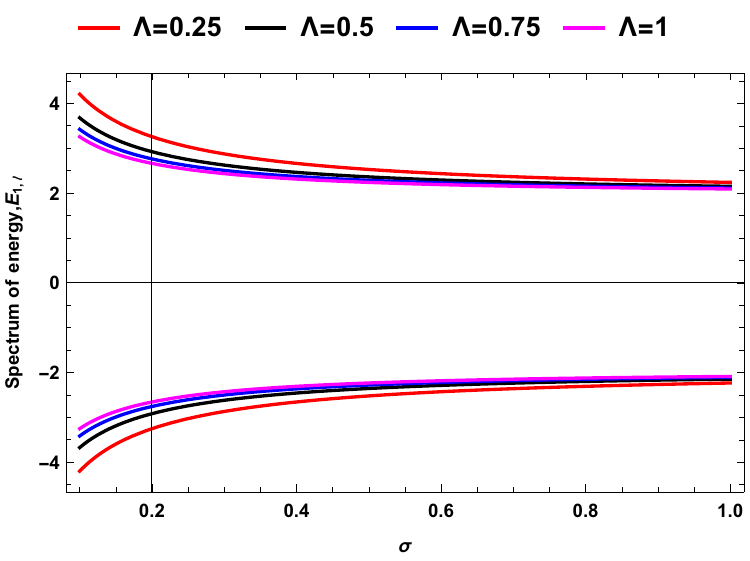}}\quad\quad\quad
\subfloat[]{\centering{}\includegraphics[scale=0.5]{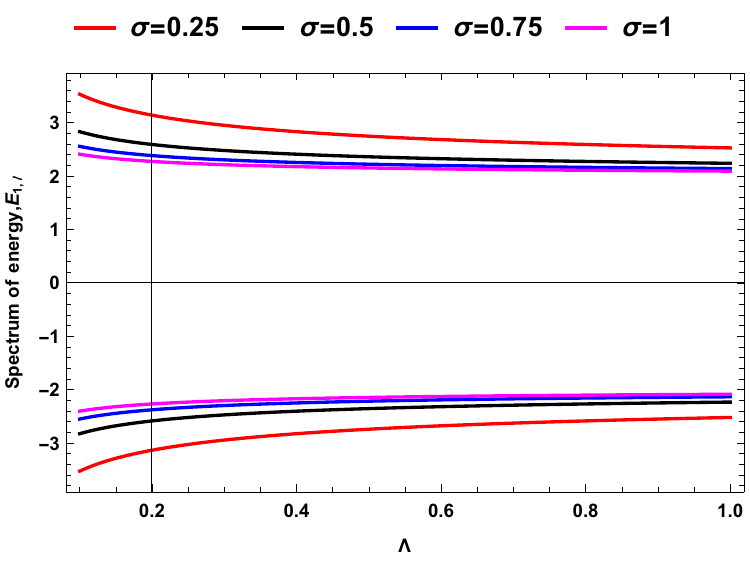}}\caption{Energy spectrum with $\sigma$ and $\Lambda$. Here $\xi=\eta=0.5$, $j=k=M=1$}
\subfloat[$\Lambda=0.5$]{\centering{}\includegraphics[scale=0.5]{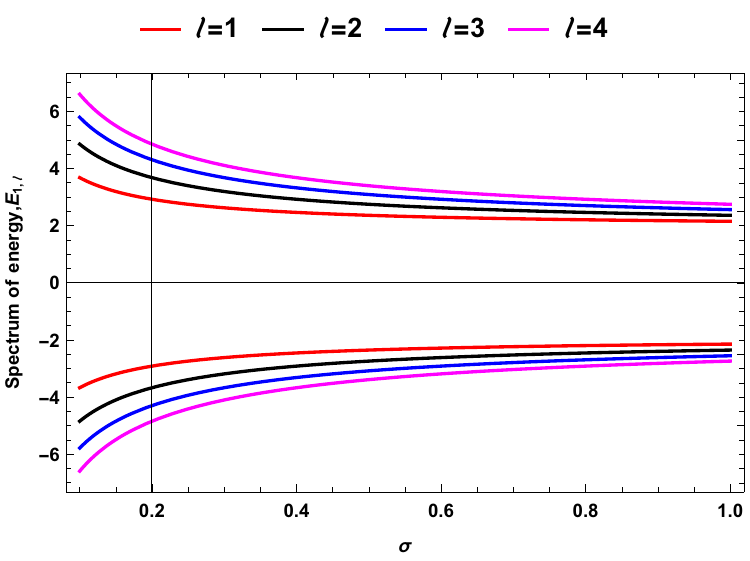}}\quad\quad\quad
\subfloat[ $\sigma=0.5$]{\centering{}\includegraphics[scale=0.5]{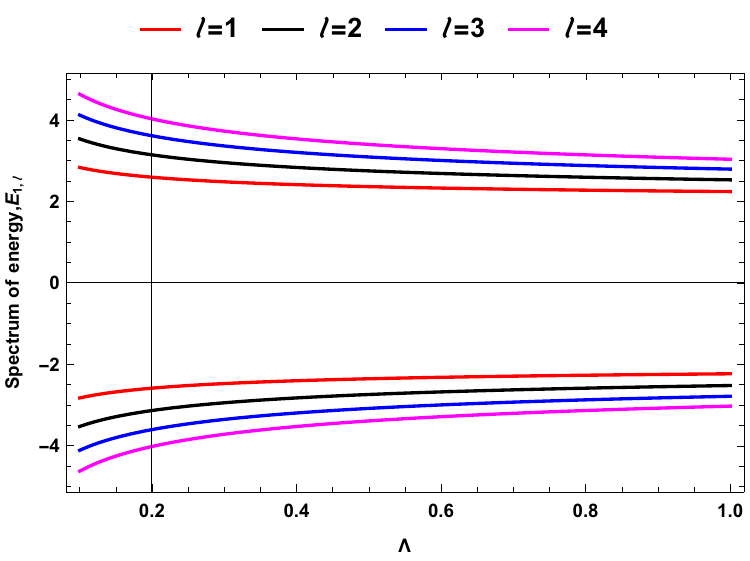}}\caption{Energy spectrum with $\sigma$ and $\Lambda$. Here, $\xi=\eta=0.5$, $j=k=M=1$}
\end{figure}
\par\end{center}

The corresponding approximate wave function is given by
\begin{equation}
\psi_{j,\ell}=\mathcal{N}_{2}\,\rho^{\iota}\,\mathrm{e}^{-\frac{1}{2}\,\rho^{2}\,\eta-M\,\rho}\,\mathrm{HeunB}\left(2\,\iota,\frac{2\,M}{\sqrt{\eta}},\frac{M^{2}+\beta^{2}-2\,\eta\,\xi}{\eta},\frac{4\,M\,\xi}{\sqrt{\eta}},\sqrt{\eta}\,\rho\right).\label{eq:b19}
\end{equation}

The ground-state of the quantum system is defined by $j=1$. Therefore, from (\ref{eq:b17}) we have $\mathcal{D}_{2}=0$ and $\gamma=2$. But from the recurrence relation (\ref{eq:b16}), we obtain by setting $\mathcal{D}_{2}=0$ as follows:
\begin{equation}
    \mathcal{D}_{1}=\frac{2}{\varsigma+\alpha'}\,\mathcal{D}_0.\label{eq:b20}
\end{equation}
Equating equations (\ref{eq:b14}) and (\ref{eq:b20}) and simplification results
\begin{equation}
    \eta_{1,\ell}=\frac{M^2}{\Big(\frac{1}{2}+\sqrt{\xi^2+\frac{\ell^2}{2\,\Lambda\,\sigma^2}}\Big)}\,\Bigg(\xi+\frac{1}{2}+\sqrt{\xi^2+\frac{\ell^2}{2\,\Lambda\,\sigma^2}}\Bigg)\,\Bigg(\xi+\frac{3}{2}+\sqrt{\xi^2+\frac{\ell^2}{2\,\Lambda\,\sigma^2}}\Bigg).\label{eq:b21}
\end{equation}
a constraint on the potential parameter $\eta$ that gives us allowed values of the energy level and a first order polynomial of $\Omega$.

Therefore, the approximate ground state energy level is given by
\begin{eqnarray}
    E_{1,\ell}=\pm\,\Bigg[k^2+\frac{2\,M^2\,\Big(\xi+\frac{1}{2}+\sqrt{\xi^2+\frac{\ell^2}{2\,\Lambda\,\sigma^2}}\Big)\,\Big(\xi+\frac{3}{2}+\sqrt{\xi^2+\frac{\ell^2}{2\,\Lambda\,\sigma^2}}\Big)\,\Big(2+\xi+\sqrt{\xi^2+\frac{\ell^2}{2\,\Lambda\,\sigma^2}}\Big)}{\Big(\frac{1}{2}+\sqrt{\xi^2+\frac{\ell^2}{2\,\Lambda\,\sigma^2}}\Big)}\Bigg]^{1/2}.\label{eq:b22}
\end{eqnarray}
The corresponding approximate ground-state wave function is given by
\begin{equation}
    \psi_{1,\ell}=\rho^{\iota}\,\mathrm{e}^{-\frac{1}{2}\,\rho^{2}\,\eta_{1,\ell}-M\,\rho}\,\Big[\mathcal{D}_0+\mathcal{D}_1\,\sqrt{\eta}\,\rho\Big],\label{eq:b23}
\end{equation}
where
\begin{equation}
    \mathcal{D}_1=\sqrt{\frac{\Bigg(\xi+\frac{1}{2}+\sqrt{\xi^2+\frac{\ell^2}{2\,\Lambda\,\sigma^2}}\Bigg)}{\Bigg(\frac{1}{2}+\sqrt{\xi^2+\frac{\ell^2}{2\,\Lambda\,\sigma^2}}\Bigg)\,\Bigg(\xi+\frac{3}{2}+\sqrt{\xi^2+\frac{\ell^2}{2\,\Lambda\,\sigma^2}}\Bigg)}}.\label{eq:b24}
\end{equation}

Equation (\ref{eq:b22}) is the relativistic approximate ground state energy level and equations (\ref{eq:b23})--(\ref{eq:b24}) is the corresponding approximate ground state wave function of a relativistic scalar particle in the background of Bonnor-Melvin magnetic universe in the presence of a Cornell-type scalar potential. Following the similar procedure, we one find excited states energy levels and the wave function of the quantum system defined by the mode $j \geq 2$. We see that the ground-state energy spectrum and the wave function is influenced by the topology parameter $\sigma$ of the geometry, the cosmological constant $\Lambda$, and the potential parameter $\xi$ of Coulomb-type potential. We also see that approximate ground-state energy level $E_{1,\ell}$ for scalar particle-antiparticles are equally spaced on either side about $E=0$ for a fixed quantum number $\ell$ and potential values $\xi$, indicates that energies are same both for particles-antiparticles in the presence of Cornell-type scalar potential.

\section{Conclusions }

In our investigation, we focused into the relativistic quantum dynamics of scalar particles embedded within the framework of a magnetic universe, taking into account the presence of a cosmological constant. Our focal point was the Bonnor-Melvin magnetic universe, where the strength of the magnetic field is determined by the topological parameter $\sigma$ and the cosmological constant $\Lambda$. Additionally, we introduced a static non-electromagnetic scalar potential denoted as $S$, introducing a modification to the mass term $M \to (M+S)$ within the Klein-Gordon equation. We derived the radial wave equation for this scalar potential $S$ in the magnetic universe background, opting for a linear confining potential. The resulting radial equation took the form of a Biconfluent Heun second-order differential equation. Utilizing a power series solution method, we obtained approximate solution for eigenvalues, such as the ground state energy level, and the corresponding wave function for the scalar particles. Importantly, we illustrated that the eigenvalue solution is influenced by the topology of the geometry represented by $\sigma$ and the cosmological constant $\Lambda$, leading to notable modifications. 

Next, we selected a Cornell-type scalar potential, and following the established procedure, we presented approximate ground state energy level and the radial wave function of the quantum system. In this case, we observed that the eigenvalue solution is not only influenced by the parameters $(\sigma, \Lambda)$ but also by the potential parameters $(\eta, \xi)$. This underscores the intricate interplay between the magnetic universe's topology, the cosmological constant, and the scalar potential, showcasing the behaviour of scalar particles in this magnetic universe space-time background.

The strength of the magnetic field $H(\rho)$, as expressed in Eq. (\ref{eq:4}), is notably influenced by two key parameters: the cosmological constant $\Lambda$, and the topological parameter $\sigma$ which introduces an angular deficit of the angular coordinate.In Figure 1, we illustrate this dependency by showcasing the field for various values of these parameters. 

As previously discussed, the dynamics of spin-0 scalar particles/fields in the presence of static scalar potentials are significantly impacted by these parameters. Consequently, alterations in the behavior of spin-0 scalar particles/fields within the considered curved space-time are induced by these parameters, which concurrently affect the strength of the magnetic field.

We have generated several figures to depict the approximate energy spectrum (\ref{eq:aa19}) and (\ref{eq:b22}) in relation to the topology parameter $\sigma$ and the cosmological constant $\Lambda$. In Figure 2(a), it is evident that as the values of the cosmological constant $\Lambda$ increase, while maintaining $\ell=1=k=M$ constant in a suitable chosen system of units where $\hbar=c=G=1$, the approximate energy levels decrease. The downward shift is more pronounced with increasing values of the parameter $\sigma$. A similar trend is also observable in Figure 2(b) as the values of the topology parameter $\sigma$ increase. Moving on to Figure 3(a), for a fixed value of the cosmological constant $\Lambda=0.5$, the decreasing energy levels shift upward with an increase in the angular momentum quantum number $\ell$. A parallel observation can be made in Figure 3(b) for a fixed value of $\sigma=0.5$. The trends depicted in Figures 4 and 5 can be explained in a similar manner. It is important to note that these figures provide valuable insights into the interplay between the cosmological constant, topology parameter, and other relevant quantum parameters in the quantum systems under investigations. The observed shifts in energy levels offer a deeper understanding of how these factors influence the dynamics of the scalar particles within the magnetic universe.

In the interpretation of particle and anti-particle states within the spin-0 system, it's crucial to clarify their association with positive and negative energy solutions to the Klein-Gordon equation within the framework of quantum field theory. These solutions correspond to excitation's of the scalar particles/fields, which can manifest as either particles or anti-particles depending on their energy.

In our study, we provided an elaboration on how these fundamental concepts relate to the solutions derived here and their implications for the behavior of scalar particles/fields within magnetic universes. This elucidation aids in understanding the nuanced dynamics of the system and its implications for particle physics in curved space-time.

\section*{Conflict of Interest}

There is no conflict of interests in this paper.

\section*{Data Availability Statement}

No new data are generated or analysed during this study.

\section*{Acknowledgments}

We would like to thank the anonymous referee's for their positive comments, critics and helpful suggestions to improved the manuscript. F.A. acknowledges the Inter University Centre for Astronomy and Astrophysics (IUCAA), Pune, India for granting visiting associateship.

\section*{Funding Statement}

There is no funding agency associated with this manuscript.

\end{document}